\documentclass[dvipsnames, sigconf, anonymous=false]{acmart}

\usepackage{xcolor}
\usepackage{booktabs}
\usepackage{listings}
\usepackage{enumitem}
\usepackage{multirow}
\usepackage{array}
\usepackage{protobuf-lang} 

\newcommand{\an}[1]{$^{_{_{^{^{#1}}}}}$}

\newcommand{\anserini}{Anserini}
\newcommand{\jass}{JASSv2}
\newcommand{\pisa}{PISA}
\newcommand{\terrier}{Terrier}
\newcommand{\olddog}{OldDog}
\newcommand{\lucene}{Lucene}

\usepackage{xspace}
\newcommand{\ciff}{\textsc{Ciff}\xspace}

\newcommand{\robust}{\texttt{Robust04}}
\newcommand{\clueweb}{\texttt{ClueWeb12B}}

\setcopyright{none}
\renewcommand\footnotetextcopyrightpermission[1]{} 

\begin{document}

\title{Supporting Interoperability Between Open-Source Search Engines with the Common Index File Format}

\author{Jimmy Lin,\an{1} Joel Mackenzie,\an{2} Chris Kamphuis,\an{3} Craig Macdonald,\an{4} Antonio Mallia,\an{5}}
\author{Micha\l{} Siedlaczek,\an{5} Andrew Trotman,\an{6} Arjen de Vries\an{3}}

\affiliation{\vspace{0.1cm}
$^1$ University of Waterloo \qquad $^2$ The University of Melbourne \qquad $^3$ Radboud University \\
$^4$ University of Glasgow \qquad $^5$ New York University \qquad $^6$ University of Otago
}

\renewcommand{\shortauthors}{Lin et al.}
\pagestyle{empty}

\begin{abstract}
There exists a natural tension between encouraging a diverse ecosystem of open-source search engines and supporting fair, replicable comparisons across those systems.
To balance these two goals, we examine two approaches to providing interoperability between the inverted indexes of several systems.
The first takes advantage of internal abstractions around index structures and building wrappers that allow one system to directly read the indexes of another.
The second involves sharing indexes across systems via a data exchange specification that we have developed, called the Common Index File Format (\ciff).
We demonstrate the first approach with the Java systems {\anserini} and {\terrier}, and the second approach with {\anserini}, {\jass}, {\olddog}, {\pisa}, and {\terrier}.
Together, these systems provide a wide range of implementations and features, with different research goals.
Overall, we recommend \ciff as a low-effort approach to support independent innovation while enabling the types of fair evaluations that are critical for driving the field forward.
\end{abstract}

\maketitle

\section{Introduction}

Academic information retrieval researchers often share their innovations in open-source search engines, a tradition that dates back to the SMART system in the mid 1980s~\cite{Buckley85}.
Today, there exists a vibrant ecosystem of IR toolkits capturing a variety of ranking models, query evaluation techniques, and other research innovations.
Yet, as several replicability and reproducibility efforts have shown, it is often difficult to compare different systems in a fair manner, both in terms of retrieval effectiveness and query evaluation efficiency, on standard test collections~\cite{Lin_etal_ECIR2016,Clancy_etal_OSIRRC2019_overview}.
In terms of effectiveness, many mundane details such as the stemmer, stopwords list, and other difficult-to-document implementation choices matter a great deal, often having a greater impact than more substantive differences such as ranking models.
These issues also affect efficiency-focused studies---for example, the presence or absence of stopwords alters skipping behavior during postings traversal.

On the one hand, a vibrant intellectual community demands diversity in terms of the tools available to researchers.
On the other hand, the ability to conduct meaningful evaluations across systems is critical to driving progress.
How can we meaningfully balance these two desiderata?
Despite explorations of alternative formulations of keyword search~\cite{Boytsov_etal_CIKM2016}, inverted indexes and associated data structures remain at the heart of nearly all IR systems today.
Thus, if we are able to devise a mechanism for different search engines to share index structures, this would represent substantial progress towards achieving our aforementioned goals. 

In principle, there are two ways such sharing can be accomplished:\
Since most search engine implementations have internal abstractions of index structures---providing support for basic operations such as postings lookup and traversal---it may be possible for one search engine to directly read the index structures created by another through an intermediate adaptor or wrapper.
Alternatively, we could define a data exchange format through which one system exports its index, to be imported by another system.
For expository convenience, we refer to the first as the ``wrapper'' approach and the second as the ``data exchange'' approach. 

There are advantages and disadvantages to both approaches.
The wrapper approach is only possible if the search engine implementations provide the necessary internal abstractions and that their definitions are (reasonably) aligned; feasibility is further constrained by technical practicalities.
For example, interoperability might be possible between two JVM-based systems, but bridging a Java and a C++ implementation might be too onerous.
Furthermore, this approach requires $n \times (n-1)$ distinct wrappers to support interoperability between $n$ systems, as every system would need to wrap the index structures of every other system.
A final disadvantage is the overhead involved in these wrappers, which might make fair efficiency-focused evaluations difficult to conduct.

A data exchange approach presents a different set of tradeoffs.
As such structures are not meant to be operated on directly, each system would need to read the data and rewrite the indexes into the system's native representation, necessitating an extra step to enable interoperability.
In order for the format to be general and robust, it is likely to be more verbose than each engine's native encoding, and thus this approach has the disadvantage of requiring researchers to distribute (across the network) large files that may be unwieldy to manipulate.
On the plus side, though, this approach avoids quadratic interactions, as each system would only need to write an exporter and an importer of the exchange format to support full interoperability~\cite{Crane_etal_WSDM2017}. 
Finally, data exchange incurs no performance penalty at query time, and thus can support fair efficiency evaluations.

This work demonstrates both approaches.
First, we apply the wrapper approach to bridge {\terrier} and {\anserini}, both Java-based systems.
Second, we propose a Common Index File Format (\ciff) and have built an index exporter that converts {\lucene} indexes into this format.
Additionally, we have implemented importers to take \ciff and transform the data into the native representations of four other systems ({\jass}, {\olddog}, {\pisa}, and {\terrier}), demonstrating interoperability by data exchange in practice.

After presenting experimental results using both approaches, we recommend the second and our Common Index File Format (\ciff) as the preferred method to enable rapid, decoupled independent research and exploration of ideas while enabling fair comparisons between systems that are critical to advancing the field.

\section{Experimental Setup}
\label{section:setup}

Our efforts brought together researchers who have built a number of open-source search engines (listed alphabetically by system):

\begin{itemize}[leftmargin=*]

\item {\bf \anserini}~\cite{Yang_etal_JDIQ2018} is an IR toolkit built on the popular open-source Lucene search library.

\item {\bf \jass}~\cite{trotman-jass}, written in C++, uses an impact-ordered index and processes postings Score-at-a-Time.  JASS can index TREC collections directly, but imports web collection indexes from ATIRE.

\item {\bf \pisa}~\cite{pisa19-osirrc} is an efficiency-focused search system, containing many state-of-the-art indexing and retrieval techniques. {\pisa} primarily uses document-ordered indexes and Document-at-a-Time query evaluation.

\item {\bf \olddog}~\cite{olddog-osirrc} is an IR engine built using a relational database, named after the work of \citet{Muhleisen_etal_SIGIR2014}.
Its design supports rapid prototyping through formulation of different SQL queries.

\item {\bf \terrier}~\cite{macdonald2012puppy} is an IR toolkit, first released in 2004. It is written in Java, and supports a large number of TREC collections and retrieval approaches, from BM25 to learning-to-rank.

\end{itemize}

\noindent For our experiments, we used the following two test collections:

\begin{itemize}[leftmargin=*]

\item {\bf \robust:} TREC Disks 4 \& 5, excluding Congressional Record, with topics and relevance judgments from the {\it ad hoc} task at TREC-6 through TREC-8 as well as the Robust Tracks from TREC 2003 and 2004 (topics 301--450, 601-700).

\item {\bf \clueweb:} The ClueWeb12-B13 web crawl from Carnegie Mellon University, with topics and relevance judgments from the TREC 2013 and 2014 Web Tracks   (topics 201--300).

\end{itemize}

\noindent The first is perhaps the most widely used test collection in IR, and thus readily provides different points of comparisons with the literature.
The goal of using \clueweb~is to demonstrate the scalability of our approach---to provide a sense of how large \ciff can get, and to confirm that these data can still be manipulated on modern hardware with reasonable ease.

\section{Wrappers}

As an example of the wrapper approach, we describe how interoperability between {\terrier} and {\anserini}, both Java-based systems, is achieved by wrapping the {\lucene} indexes generated by {\anserini} in {\terrier} APIs, such that {\terrier} can directly traverse {\lucene} postings for query evaluation.

The Terrier wrapper\footnote{\url{https://github.com/cmacdonald/terrier-lucene}} we have implemented for the Lucene \texttt{IndexReader} API allows a Terrier postings list iterator to directly call the underlying Lucene methods; the change is entirely transparent to Terrier.
This works well for simple frequency-based and positional representations, but we did not implement index fields due to differences in how they are defined in the two systems.

\begin{table}[t]
\centering
\begin{tabular}{l cc }
\toprule
System & AP & P@30 \\
\midrule
Anserini (BM25) & 0.2531 & 0.3102\\
Anserini (BM25+RM3) & 0.2903 & 0.3365\\
Anserini (BM25+Axiomatic QE) & 0.2896 & 0.3333\\
\midrule
Terrier (BM25) & 0.2530 & 0.3106\\
Terrier (BM25+Bo1 QE)& 0.2931 & 0.3406 \\
Terrier (BM25+RM3)& 0.2945 & 0.3371 \\
\midrule
Terrier-Lucene (BM25) & 0.2524 & 0.3091\\
Terrier-Lucene (BM25+Bo1 QE) & 0.2890 & 0.3356\\
Terrier-Lucene (BM25+RM3) & 0.2887 & 0.3284 \\
\bottomrule
\end{tabular}
\vspace{0.2cm}
\caption{Comparison of Anserini, Terrier, and the Terrier wrapper for Anserini's Lucene indexes (Terrier-Lucene) on {\robust}.}
\label{tab:wrapping}
\vspace{-0.75cm}
\end{table} 

Results on {\robust} are shown in Table~\ref{tab:wrapping}, where it is now possible to compare different query expansion methods using essentially the same index.
We note that differences in BM25 effectiveness are very small, while the various query expansion methods have at most 2\% AP difference.

Despite the feasibility of the wrapper approach in this case, we felt that the efforts involved were too substantial to be scaled to more systems.
In particular, since Terrier and Anserini were both implemented in Java, API-level integration was not too onerous.
However, bridging either with, for example, a system implemented in C++ such as PISA or JASSv2, would involve substantially more effort.
This motivated us to explore the data exchange approach more thoroughly.

\section{Common Index File Format}

Our second approach to supporting interoperability among different search engines is to define a data exchange specification that we call the Common Index File Format (\ciff) whereby systems can share their inverted indexes and other associated data structures that are required for ranking.
Critically, we intend for this to be an {\it exchange} format and not an {\it operational} one---that is, we expect each system to read \ciff and transform the contents into the system's own internal representation.

Before describing the format, we first discuss some design goals and non-goals.
We intend for \ciff to cover structures that are common to all search engines based on inverted indexes, a sort of ``lowest common denominator''.
The format must be language agnostic and easy to read.
Speed of reading/writing this format as well as compactness are {\it not} important concerns, since the format is not meant to be computed over;
thus, we specifically eschew exotic compression schemes that may result in smaller output sizes at the cost of decoding complexity.

At a high level, \ciff defines a specification for serializing postings lists and other associated data structures necessary for search engines.
Put into practice, the simplest workable exchange format could be based on plain text files.
Postings lists have regular, repeating structure, and in principle, it would be possible to define a delimited text format for capturing these structures.
However, we decided against this approach for several reasons:\ In such a scheme, metadata such as the semantics of the delimiters would need to be documented separately, and thus easily ``lost''.
Additionally, there is no easy way to enforce the integrity and validity of a particular export---unless we explicitly build in error checks, in which case the format becomes even more complicated, further exacerbating the above challenge.

\begin{figure}[t]
\small
\begin{tabular}{c}
\begin{lstlisting}
message Header {
  int32 version = 1;  
  int32 num_postings_lists = 2;  
  int32 num_docs = 3;  
  int32 total_postings_lists = 4;
  int32 total_docs = 5;
  int64 total_terms_in_collection = 6;
  double average_doclength = 7;
  string description = 8;
}

message Posting {
  int32 docid = 1;
  int32 tf = 2;
}

message PostingsList {
  string term = 1;  
  int64 df = 2;  
  int64 cf = 3; 
  repeated Posting postings = 4;
}

message DocRecord {
  int32 docid = 1; 
  string collection_docid = 2; 
  int32 doclength = 3;  
}
\end{lstlisting}
\end{tabular}
\vspace{-0.25cm}
\caption{Protobuf definitions of messages in \ciff.}
\label{fig:def}
\vspace{-0.5cm}
\end{figure}

Ultimately, we decided to use Protocol Buffers (protobufs) for serialization.
Protocol Buffers\footnote{\url{https://developers.google.com/protocol-buffers}} are a language-neutral, platform-neutral extensible mechanism for serializing structured data that is widely deployed in industry.
Protobufs share some similarities with C \texttt{struct}s in providing a language to define abstract data types that can be arbitrarily nested and repeated to represent lists.
Fields can either be referenced by name (e.g., \texttt{docid} in a \texttt{Posting}) or a numeric id.
The protobuf specification restricts types to those found on nearly all platforms (e.g., 32-bit integers) and from a definition, the protobuf compiler can automatically generate code for reading and writing data in the specified format, supporting a multitude of languages and platforms.

The protobuf messages defined in \ciff are shown in Figure~\ref{fig:def}.
In terms of these definitions, a \ciff export is comprised of a single, possibly compressed, file with a sequence of delimited protobuf messages, exactly as follows:

\begin{itemize}[leftmargin=*]

\item a \texttt{Header} message, followed by 
\item exactly the number of \texttt{PostingsList} messages specified in the \texttt{num\_postings\_lists} field of the \texttt{Header}, followed by 
\item exactly the number of \texttt{DocRecord} messages specified in the \texttt{num\_docs} field of the \texttt{Header}.

\end{itemize}

\noindent A \ciff export begins with a {\tt{Header}} that captures metadata such as versioning information, global index statistics, and a description of how the export was generated.
The \texttt{num\_postings\_lists} field specifies the number of postings lists that are included in a particular export, which allows \ciff to support the use case of including only postings lists that correspond to a particular set of evaluation topics.
Naturally, such a setting yields an export that is far smaller than the export of a complete index.
A {\tt{PostingList}} contains the term, its document frequency, its collection frequency, and a number of individual \texttt{Posting} messages equal to the document frequency.
Following standard conventions, the \texttt{docid} is encoded as gaps.
A \ciff export ends with document-specific information that is captured by a series of {\tt{DocRecord}} messages, which contain the integer docid (referenced in the postings lists), the external collection docid (a string), and the length of the document.

The complete \ciff specification, including a reference implementation that generates (and reads) \ciff
exports from Lucene indexes built by Anserini, is open-source and available in our GitHub  repository.\footnote{\url{http://ciff.osirrc.io/}}
We have also implemented importers for all the other systems described in Section~\ref{section:setup}; links to code can also be found in our repository.
The complete Anserini Lucene \ciff exports of the {\robust} and {\clueweb} indexes used in our experiments are 162 MiB and 25 GiB, respectively, as gzipped files.
Exports that contain only the query terms are 17 MiB and 1.3 GiB (compressed), for the same two collections, respectively.
Links to all these exports can also be found in our repository.
We see that, even for reasonably large web collections that are commonly-used in information retrieval research, \ciff exports are modest in size for modern hardware, both to ship across the network and to manipulate on disk.

\subsection{Case Study:\ BM25 Variants}

With \ciff, it is possible to conduct meaningful evaluations of ranking models from diverse systems that completely factor out the effects of different document processing pipelines (i.e., document cleaning regimes, tokenization, stopwords, etc.).
We illustrate with a simple case study examining ``BM25''.

One major finding from previous replicability studies~\cite{Lin_etal_ECIR2016,Clancy_etal_OSIRRC2019_overview} is that systems purporting to implement BM25 can exhibit large effectiveness differences on standard test collections.
This is due to a combination of two factors:\ First, systems have different document processing pipelines; details like data cleaning make a big difference, but are relatively uninteresting to researchers.
Second, ``BM25'' actually encompasses a large number of variants.
However, Trotman et al.~\cite{Trotman_etal_ADCS2014} and Kamphuis et al.~\cite{Kamphuis_etal_ECIR2020} found that such differences are unlikely to be statistically significant.
In both cases, this conclusion was arrived at by the authors implementing {\it all} the variants in {\it the same search engine} to support the comparisons.
Needless to say, this is a time-consuming task, and not scalable in the general case, where we would like to compare {\it arbitrary} ranking functions from {\it any} search engine.
This is exactly where \ciff comes in:\ with our exchange format, it is possible to conduct fair evaluations of ranking effectiveness on {\it different} systems.

\begin{table}[t]
\vspace{0.2cm}
\centering
\begin{tabular}{l cc cc}
\toprule
\multirow{2}{*}{System}
& \multicolumn{2}{c}{\robust} & \multicolumn{2}{c}{\clueweb}   \\
\cmidrule(lr){2-3}  \cmidrule(lr){4-5}  
& AP & P@30 & NDCG & ERR \\
\midrule
\multicolumn{4}{l}{{\it Native Document Processing}} \\[0.5ex]
JASSv2 & 0.2570 & 0.3157 & 0.1132 & 0.0809 \\
PISA & 0.2543 & 0.3139 & 0.1169 & 0.0845 \\
Terrier &0.2530 & 0.3106 & 0.1308 & 0.0978   \\
\midrule
Anserini  & 0.2531 & 0.3102 & 0.1340 & 0.0970\\
\midrule
\multicolumn{4}{l}{{\it Common Index File Format}} \\[0.5ex]
JASSv2 & 0.2524 & 0.3096 & 0.1311 & 0.0937\\
PISA   & 0.2519 & 0.3083 & 0.1345 & 0.0971\\
OldDog-A & 0.2531 & 0.3102 & 0.1345 & 0.0971 \\
OldDog-L & 0.2530 & 0.3102 & 0.1345 & 0.0971 \\
Terrier & 0.2524 & 0.3091 & 0.1321 & 0.0956 \\
\bottomrule
\end{tabular}
\vspace{0.2cm}
\caption{Comparison of BM25 variants. }
\label{tab:bm25}
\vspace{-0.7cm}
\end{table} 

\begin{figure*}[t]
\centering\includegraphics[width=0.45\linewidth]{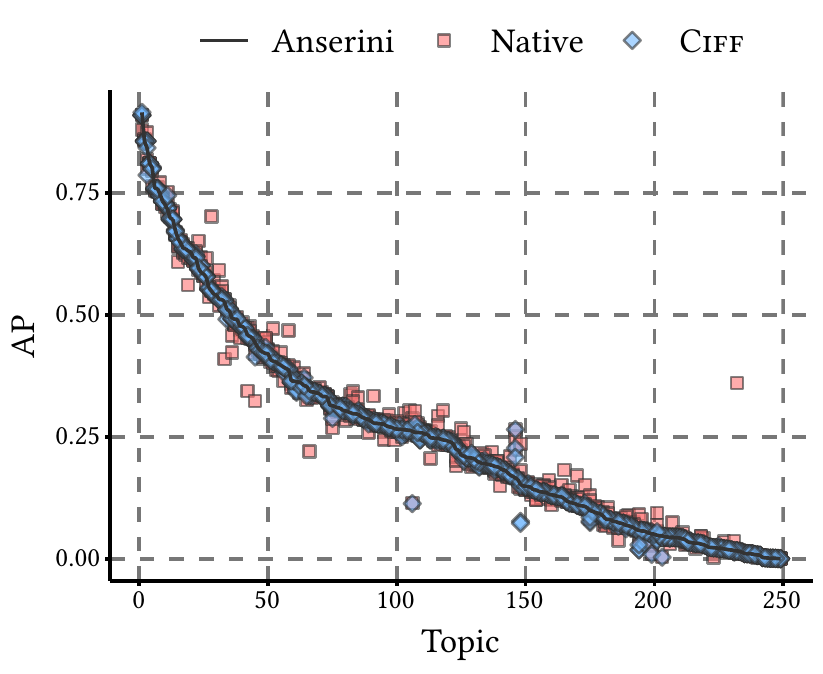}
\centering\includegraphics[width=0.45\linewidth]{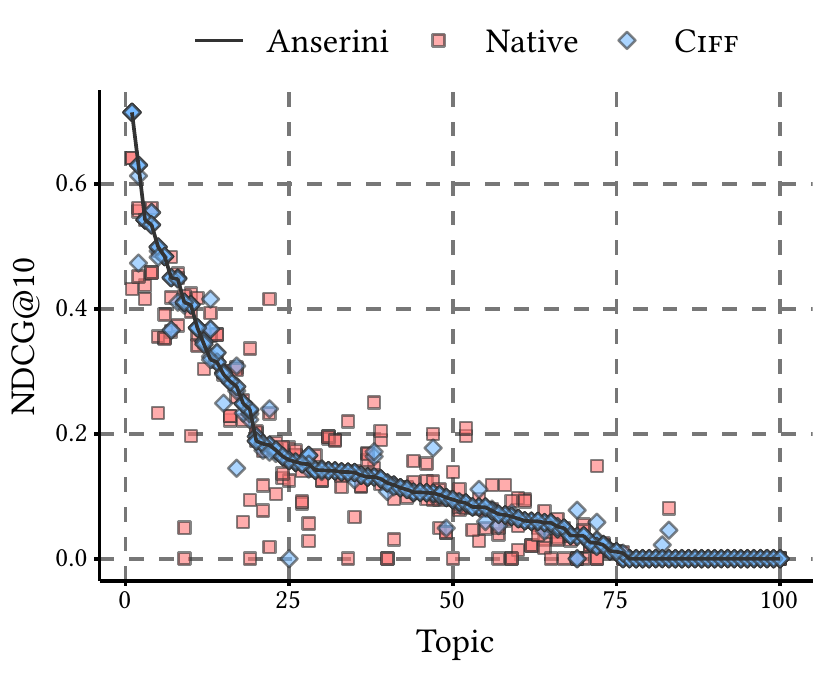}
\vspace{-0.35cm}
\caption{Per-topic scores for all systems, sorted in descending order of the metric ($y$-axis), based on Anserini's scores:\ {\robust} on left and {\clueweb} on right.}
\vspace{-0.1cm}
\label{figure:robust}
\end{figure*}

To illustrate, we present a simple, multi-system study of BM25 variants.
For {\robust} and {\clueweb}, we exported the Lucene indexes generated by Anserini into \ciff{} (see previous section), which is then imported by all the remaining systems.
We evaluated each system's BM25 ranking using standard metrics:\ AP (at rank 1000) and P@30 for \robust{}, NDCG@10 and ERR@10 for \clueweb{}.
In all cases we set $k_1=0.9$ and $b=0.4$, per the recommendations of~\citet{Trotman_etal_2012}.
These results are shown in Table~\ref{tab:bm25}.
In the top block of the table, we present figures from each system's ``native'' document processing pipeline to provide points of reference.
Note that since Anserini is built directly on Lucene, its \ciff and ``native'' results are identical.
For OldDog, we report both ATIRE BM25 (OldDog-A) and Lucene BM25 (OldDog-L).

There are three sources of differences in systems' rankings:\ (1) implementation of the document processing pipeline, (2) variants of the BM25 scoring function (including different parameter settings, quantization effects when computing impact scores, etc.), and (3) tie-breaking effects.
With \ciff, we have eliminated the first effect.
The third effect has been characterized in previous work~\cite{Lin_Yang_SIGIR2019,Yang_etal_AIRS2016} and is mitigated here because \ciff ensures that documents are consistently ordered across all systems.
Thus, this experiment allows us to isolate the effects of BM25 variants, although we must still manually ensure that every system uses the same parameter settings.
In short, we have replicated previous replicability studies~\cite{Trotman_etal_ADCS2014,Kamphuis_etal_ECIR2020}, but in a manner that supports {\it cross-system} comparisons.

From Table~\ref{tab:bm25}, we see that effectiveness differences between the various systems with native document processing are larger than with \ciff.
This effect is particularly noticeable with \clueweb:\ on web documents, document processing (e.g., cleaning of HTML) has a much larger impact on effectiveness compared to {\robust}, which comprises relatively clean SGML documents.

We conducted a Tukey's HSD (honestly significant difference) test for all the ``native'' systems as a group and with \ciff as a group:\ none of the differences are statistically significant, for both {\robust} and {\clueweb}.
Nevertheless, if we examine per-topic scores, the differences between each system's native document processing pipeline and \ciff become much more prominent.
Consider Figure~\ref{figure:robust} (left), which plots the per-topic AP scores for {\anserini} on {\robust}, in decreasing order of effectiveness.
We have overlaid the scores for the corresponding topics from all systems for both the native and \ciff conditions.
Clearly, we see that {\ciff} reduces most of the per-topic effectiveness differences between systems. 
This experiment was repeated on the {\clueweb{}} collection, using NDCG@10 as the metric; results are shown in Figure~\ref{figure:robust} (right).
Once again, although there remain differences between systems' scores under \ciff, the conflating issue of document processing has been eliminated, thereby allowing researchers to more meaningfully characterize effectiveness.

Our simple case study demonstrates how \ciff supports meaningful cross-system comparisons, albeit on a simple, well-worn example.
However, our approach can be easily extended to evaluations of different ranking models, candidate generation techniques in multi-stage ranking pipelines, performance comparisons of query latency, and beyond.

\section{Conclusions}

We envision \ciff to be an ongoing, open, and community-driven effort that allows researchers to independently pursue their own lines of inquiry while supporting fair and meaningful evaluations.
Additional contributions are most welcome!
As our efforts gain traction, we envision future research papers adopting ``standard'' \ciff exports in their experiments---this would have the dual benefit of standardizing empirical methodology and more clearly highlighting the impact of proposed innovations.

\section*{Acknowledgments}

This research was supported in part by the Natural Sciences and Engineering Research Council (NSERC) of Canada, Compute Ontario and Compute Canada, the Australian Research Council (ARC) Discovery Grant DP170102231, the US National Science Foundation (IIS-1718680), and research program Commit2Data with project number 628.011.001 financed by the Dutch Research Council (NWO).

\renewcommand{\bibsep}{2.5pt}
\bibliographystyle{ACM-Reference-Format}
\bibliography{common-index}

\end{document}